\documentclass[12pt]{revtex4}

\def\figurename{Figure}
\usepackage{amsmath}
\usepackage{rotating}
\usepackage{graphicx}
\usepackage{lscape}
\usepackage[utf8]{inputenc}
\usepackage{amsfonts}
\usepackage{amssymb}
\usepackage{amsthm}
\usepackage[T1]{fontenc}
\usepackage[spanish]{babel}
\usepackage[active]{srcltx}
\usepackage{enumerate}
\usepackage{graphicx}
\usepackage{epsfig}
\usepackage{pb-diagram}
\usepackage{tensor}
\usepackage{bm}

\begin{document}
\newcommand{\nl}{{\cal N}}
\newcommand{\cl}{\Phi}
\def\binom#1#2{{#1\choose #2}}
\newcommand{\sas}{ $\hat{S}^{2}$ and $\hat{S}_{z}$ }
\newcommand{\hu}{{\cal A}}
\newcommand{\dm}{{\cal D}}
\newcommand{\idn}{{\cal I}}
\newcommand{\lo}{\Lambda\Omega}

\renewcommand{\figurename}{\textbf{\textit{Figure}}}




\newtheorem{corollary}{Corollary}
\newtheorem{lemma}{Lemma}
\newtheorem{proposition}{Proposition}
\newtheorem{theorem}{Theorem}


\newenvironment{dem}[1][Proof]{\begin{proof}[{\bf #1}]}{\end{proof}}


\theoremstyle{definition}

\newtheorem{axiom}{Axioma}[section]
\newtheorem{definition}{Definición}[section]
\newtheorem{example}{Ejemplo}[section]
\newtheorem{remark}{Observación}[section]
\newtheorem{exercise}{Ejercicio}[section]
{\swapnumbers\newtheorem{exercisestar}[exercise]{(*)}}


\newcommand{\C}{\ensuremath{\mathbb{C}}}
\newcommand{\N}{\ensuremath{\mathbb{N}}}
\newcommand{\Q}{\ensuremath{\mathbb{Q}}}
\newcommand{\R}{\ensuremath{\mathbb{R}}}
\newcommand{\T}{\ensuremath{\mathbb{T}}}
\newcommand{\Z}{\ensuremath{\mathbb{Z}}}


\newcommand{\abs}[1]{\left\vert #1\right\vert}
\newcommand{\bra}[1]{\left\langle #1\right\vert}
\renewcommand{\dim}[1]{\mathrm{dim}\left( #1\right)}
\newcommand{\set}[1]{\left\{ #1\right\}}
\renewcommand{\ker}[1]{\mathrm{Ker}\pa{#1}}
\newcommand{\ket}[1]{\left\vert #1\right\rangle}
\newcommand{\norm}[1]{\left\|#1\right\|}
\newcommand{\pa}[1]{\left(#1\right)}
\newcommand{\pro}[2]{\left\langle#1|#2\right\rangle}
\newcommand{\proo}[3]{\left\langle#1\left|#2\right|#3\right\rangle}
\newcommand{\ran}[1]{\mathrm{Ran}\pa{#1}}
\newcommand{\tr}[1]{\mathop{\mathrm{Tr}}\pa{#1}}

\newcommand{\Rn}[1][\mathcal{N}]{\mathbb{R}^{#1}}
\newcommand{\Cn}[1][\mathcal{N}]{\C^{#1}}
\newcommand{\Sph}[1][\mathcal{N}]{\mathrm{S}^{#1}}

\newcommand{\codim}[1]{\mathrm{codim}\left( #1\right)}
\newcommand{\diag}{\mathop{\mathrm{diag}}}
\newcommand{\diam}{\mathop{\mathrm{diam}}}
\newcommand{\id}{\mathop{\mathrm{id}}}
\newcommand{\inte}{\mathop{\mathrm{int}}}
\newcommand{\inc}{\mathop{\mathrm{\iota}}}
\newcommand{\sop}{\mathop{\mathrm{sop}}}

\newcommand{\GL}[2][R]{\mathrm{GL}\pa{\mathbb{#1},#2}}
\newcommand{\GLp}[2][R]{\mathrm{GL}_{+}\pa{\mathbb{#1},#2}}
\newcommand{\SL}[2][R]{\mathrm{SL}\pa{\mathbb{#1},#2}}
\newcommand{\Or}[1][\mathcal{N}]{\mathrm{O}\pa{#1}}
\newcommand{\SO}[1][\mathcal{N}]{\mathrm{SO}\pa{#1}}
\newcommand{\Un}[1][\mathcal{N}]{\mathrm{U}\pa{#1}}
\newcommand{\SU}[1][\mathcal{N}]{\mathrm{SU}\pa{#1}}
\newcommand{\Up}[2][R]{\mathrm{Up}\pa{\mathbb{#1},#2}}
\newcommand{\Upp}[2][R]{\mathrm{Up}_{+}\pa{\mathbb{#1},#2}}
\newcommand{\her}[1]{\mathrm{her}\pa{#1}}
\newcommand{\hil}{\mathsf{H}}

\renewcommand{\Re}[1]{\mathrm{Re}\left( #1\right)}
\renewcommand{\Im}[1]{\mathrm{Im}\left( #1\right)}



\vskip -10mm

\vskip -10mm

{\Large \noindent {\bf Quantum States of Physical Domains in
Molecular  \\ Systems: A three-state model approach}}

\vskip 5mm

\noindent {\bf Roberto C. Bochicchio}$^{*,1,2}$, {\bf Boris
Maul\'en}$^{3}$

\vskip 2mm

{\small

 \noindent $^{1}$ {\it Universidad de Buenos Aires, Facultad
de Ciencias Exactas y Naturales, Departamento de F\'{\i}sica, Ciudad
Universitaria, 1428, Buenos Aires, Argentina}

\noindent $^{2}$ CONICET - {\it Universidad de Buenos Aires,
Instituto de F\'{\i}sica de Buenos Aires (IFIBA) Ciudad
Universitaria, 1428, Buenos Aires, Argentina}

\noindent $^{3}$ {\it Departamento de Ciencias Qu\'{\i}micas,
Facultad de Ciencias Exactas, Universidad Andres Bello, Santiago,
Chile}

}

\vskip 5mm



\noindent The physical regions (domains or basins) within the
molecular structure are open systems that exchange charge between
them and consequently house a fractional number of electrons (net
charge). The natural framework describing the quantum states for
these domains is the density matrix (DM) in its grand-canonical
version which corresponds to a convex expansion into a set of basis
states of an integer number of electrons. In this report, it is
shown that the solution for these quantities is supported by the DM
expansion into three states of different number of particles, the
neutral and two (edge) ionic states. The states and the average
number of particles in the domains (fractional occupation
population) are determined by the coefficients of the expansion in
terms of the fundamental transference magnitudes revealing the
donor/acceptor character of the domains by which the quantum
accessible states are discussed.

\vskip 50mm

{\small \noindent
\rule{60mm}{0.4mm} \\
\noindent $^{*}$ rboc@df.uba.ar}

\newpage
\noindent {\large {\bf 1 Introduction}}

\vskip 3mm

Isolated molecular systems (S) are composed of an interacting set of
charged particles (electrons and nuclei). They are essentially
driven by Coulomb interactions and possess a crucial experimental
feature: their energies and consequently their electronic densities
are convex functions of the number of particles when considered as a
closed system of a fixed integer number of particles
\cite{PPLB,Parr_Yang_book,Geerlings,Boch_Rial_JCP_2012}. Therefore,
it has been shown that, supported by the convexity property, they
can be handled in a two-dimensional Fock space in which Density
Matrix $D$ (DM), i.e., the quantum state, is rigorously spanned by a
two-state model by the $D$s of the neutral specie {\bf X}$^{0}$ and
only one of the ionic species {\bf X}$^{+/-}$, as has been shown in
Refs. \cite{PPLB,Boch_Rial_JCP_2012}. Also, under a small
perturbation interaction with an environment (R), namely, a
reservoir or a solvent field, the energy may vary only an amount and
then the aforementioned energy convex dependence with the number of
particles remains, i.e., it also fulfills the convexity hypothesis.
This interaction reflects the openness of the system and avoids the
energy and density derivative discontinuities to determine some
chemical descriptors like chemical potential and hardness
\cite{Wasserman,Boch_TCA2015}.

Any molecular structure chemically considered is composed of a set
of non-overlapping physical domains $\Omega$ (basins) rigorously
defined by the so-called Quantum Theory of Atoms in Molecules
(QTAIM) \cite{Bader_book,Popelier_book}. These physical domains
embedded within the molecular system can be an atom, a functional
group, or simply a moiety interacting with the other domains in the
system,  exchanging charge between them and, consequently, housing a
non-integer (fractional) number of particles (electron populations).
Therefore, the number of particles in the domain $\nl$ is required
to be a continuous variable, and thus, the domain must be treated as
an open quantum system \cite{Pendas1,Pendas2}. This kind of system,
consequently, unlike those of an integer number of particles $N$,
misses the aforementioned convexity dependence with the number of
particles. This is the essential difference between a whole
molecular system {\bf X}$^{0}$ and/or its ionic forms {\bf
X}$^{\pm}$ and a physical domain $\Omega$ inside them.

The breakdown of the convexity hypothesis and the onset of the
concept of the number of particles as a continuous variable $\nl$ of
an open system merits a larger convex expansion of $D$ of at least a
number of pure states greater than two, each one defined by their
integer number of particles $N$ \cite{PPLB,Boch_Rial_JCP_2012}. In
this way, we can build a more general quantum state for this new
scenario of non-convexity. The natural approach to perform the task
to describe the statistical electron distribution in molecules is
the theory of density matrices \cite{Blum_book} and their related
hierarchy of reduced density matrices $RDM$s \cite{McWeeny_book,
Davidson_book, Coleman_book}. So, the treatment of a fractional
occupation number and the mathematical and physical basis as well as
the solutions for this problem may be discussed within a
grand-canonical statistical distribution (GC) formalization of $D$
for few particles \cite{PPLB,Parr_Yang_book,Boch_Rial_JCP_2012}.
This subject of study makes evident the fact of the adequate use of
shared methods of statistical physics for systems of few particles.

The initial problem for the treatment is to fix the number of states
in the Fock space, namely, the dimension, in which $D$ is expanded.
For that goal, let us consider the physical image of the domains
supported by the classical view of their physicochemical behavior as
donor or acceptor entities exchanging charge with their
surroundings, namely the other domains in the molecular framework.
Following these ideas, we will show that it is possible by extending
the number of systems by only three states, i.e., the neutral one
($N$ electrons) and two ionic {\it edge} states labeled and
characterized by its maximum acceptor/donor capacity, $N \pm q$,
where $q \in \mathbb{Z}^+$ corresponds to the maximum number of
electrons that can be accepted or donated by the neutral specie,
respectively. This type of structure has been previously used within
the thermodynamical approach and Density Functional Theory (DFT)
formalism \cite{Ramon1,Ramon2} and was called $N-$centered ensemble
\cite{N_center_ensemble}.

The main goal of the present development is to seek the accessible
states $D$ for a domain and consequently to derive the expressions
for the energy in terms of the rate of transference of charge, as
the fundamental quantity for the description.  It is worth noting
that the present approach will not consider a stabilization process
due to the environmental effects R, i.e., solvation and/or the
exchange of electrons with its environment, nor the introduction of
the concept of temperature. Thus, no thermodynamic arguments are
used in the description of the system since our viewpoint is a
purely quantum-mechanical one. Also, the shape of the domain
$\Omega$, i.e., volume and its respective covering surface remain
fixed as well as the external potential given by the nuclei in the
molecule. Thus, any charge transference fraction $\nu$ in the domain
quantum state will be referred to the corresponding one in its
ground state $\nu_0$, to feature its donor or acceptor character.
The effects of changing the shape of the domain are not the subject
of this work and will be considered in forthcoming studies.

This article is organized as follows. The Second Section is devoted
to the introduction of the mathematical framework for the model and
the solutions to the problem. The Third Section stands for the
discussion and final remarks.

\vskip 10mm

\noindent {\large {\bf 2 The model}}

\vskip 3mm

Before introducing the mathematical model it is necessary to
establish more precisely the concept of physical domain in a
molecular system. To this end, we consider the aforementioned domain
as composed by a collection of atoms arranged in a chemical
functional group or an arbitrary moiety conveniently defined or even
as an individual atom, within the molecular structure. Thus, each
domain is a three-dimensional region inside the molecular frame and
corresponds to any of these units according to the particular
interest of the study. Since each domain is embedded in the whole
molecular frame, it behaves as an open quantum system, and, due to
the net charge transfer with the other interacting domains, it
houses a non-integer number of electrons. Different methodologies
allow to describe the physical domains within a molecule. Some of
them have semi-empirical roots \cite{fuzzy_atoms,elf}, and the only
one with rigorous foundations is the Quantum Theory of Atoms in
Molecules (QTAIM) \cite{Bader_book,Popelier_book}. Examples of these
domains are, for instance, $CO$ or $OH$ groups in formaldehyde and
methanol, respectively \cite{CPL_BO,CPL_V}; $Fe$ metal atom in
$[Fe(CO)_{6}]^{+2}$ cluster \cite{Lobayan_clusters}; $CC$ moiety in
ethylene. Hence each domain $\Omega$ can contain one or more atoms.

The number of electrons in the neutral state of the molecular domain
is the sum of the $k$-th atomic number $Z_{k}$ of each atom in the
$\Omega$ domain, and it is expressed by $N=Z_{\Omega}=\sum_{k\subset
\Omega} Z_{k}$. On the other hand, the anionic and cationic states
possess a number of electrons given by $Z_{\Omega}+q$ and
$Z_{\Omega}-q$, respectively, being both integer numbers (where $q$
stands for the maximum charge that the domain can accept or donate).
So, the electronic population of a region $\Omega$, $N_{\Omega}$
\cite{CPL_BO,CPL_V}, becomes defined by means of a non-integer
number $\mathcal{N}=N+\nu$, which expresses the acceptor/donor
character of the domain, where $\nu$ is a real number defined in the
interval $-q \le \nu \le q$ and corresponds to the fraction of
charge transferred to/from the domain.

It is of fundamental importance to stress the fact that for this
type of systems, i.e., the molecular domains, the energy convexity
under the variation of the number of particles can not be invoked
unlike for the whole isolated molecule, as anticipated in the
previous section \cite{PPLB,Parr_Yang_book,Boch_Rial_JCP_2012}.

The quantum states associated with the systems described above
(molecular domains) can be handled within the density matrix theory
in the GC ensemble representation of pure states with different
number of particles $M$. Thus, for a density matrix $D$ representing
a quantum (mixed) state in a molecular domain, we have

\begin{equation}
D=\sum_{M}\sum_{k} \omega_k^M \ket{\Phi_{k}^M}\bra{\Phi_k^M}=\sum_M
\sum_k \omega_k^M \hspace{0.05cm}{}^{M}\hspace{-0.1cm} D_k,
\label{D1}
\end{equation}

\noindent where ${}^{M}D_k=\ket{\Phi_k^M}\bra{\Phi_k^M}$ stands for the $k$-th
$M$-electron pure state in an antisymmetric $M$-electron Hilbert
space $\mathcal{H}_M$ (Hamiltonian eigenstates), and $\omega_k^M$
correspond to the statistical weights of a grand-canonical-like
distribution, which satisfy

\begin{equation}
\sum_M \sum_k \omega_k^M =1 \hspace{0.5cm} 0\leq \omega_k^M \leq 1.
\label{constraint_statistical_weights}
\end{equation}
Also, it is important to mention that a density matrix $D$ lives in
the Fock Space $\mathcal{F}$, i.e., $D\in \mathcal{F}$, where
$\mathcal{F}$ is built as a direct sum of $M$-particle Hilbert
spaces $\mathcal{H}_M$ \cite{Emch}

\begin{equation}
\nonumber \mathcal{F}=\bigoplus_{M=0}^{\infty}\mathcal{H}_M.
\end{equation}

Once we have established the scenario by definition of the system
and its corresponding states, we introduce a three-state model which
simplifies and, to some extent, generalizes the treatment due to the
introduction of the parameters $\nu$ and $q$. In this way, the
statistical weights $\omega^{M}_{k}$ are considered as explicit
functions of the fraction of charge $\nu$ and $q$, i.e.,
$\omega^{M}(\nu/q)$, as it will become evident hereunder. At this
point, there are two important facts worth mentioning: first, the
total charge $q$ corresponds to a physical parameter that defines
the nature of the molecular domain; and second, in this model we
only deal with electronic ground states, whereby, for now on we can
omit the label $k$ associated to the quantum number in the
statistical weights.

The basis states for the convex expansion are the ground states
$^{M}D_{0}$ of the $M$-particle systems. In this way, the quantum
state of a molecular domain $\Omega$ reads

\begin{equation}
D= \omega^{N-q}(\nu/q)\; {^{N-q}D}_{0} +\;
\omega^{N}(\nu/q)\;{^{N}D}_{0} +\; \omega^{N+q}(\nu/q)\;
{^{N+q}D}_{0}. \label{D2}
\end{equation}

\vskip 3mm \noindent Thus, the three-states model retains only three
states in Eq. (\ref{D1}), namely, the central (neutral) one having
$N$ electrons and two {\it edge} states, the cationic/anionic of at
most $\pm q$ electrons, i.e., $N \pm q$, respectively.

The solution to this problem comes from two well-established
statistical conditions, namely, the zero and first-order moments of
the electron distribution. They define algebraic relations for the
statistical weights. The zero-order moment stands for the
normalization condition for the quantum state, and it is expressed
by

\begin{subequations}
\begin{equation}
\omega^{{N-q}}(\nu/q) +\; \omega^{{N}}(\nu/q) +\;
\omega^{{N+q}}(\nu/q)\; \label{zero-order} = 1. \hspace{1.0cm}
\end{equation}

\vskip 3mm

\noindent On the other hand, the first-order moment is a condition
on the occupation (electronic population) of the domain $\Omega$,
$N_{\Omega}$, and establishes that the mean value of the number of
particles $\left\langle M \right\rangle$ given by

\begin{equation}
\nonumber \left\langle M \right\rangle =\sum_M \omega^M M,
\end{equation}

\noindent with {\small $M=N-q,N,N+q$}, must agree with $\mathcal{N}=N+\nu$,
i.e.,

\begin{equation}
\left\langle M \right\rangle=N+\nu, \label{first-order}
\end{equation}

\noindent leading thus to the relation

\vskip 3mm

\begin{equation}
\omega^{{N+q}}\; -\; \omega^{{N-q}}\; =\; \frac{\nu}{q}.
\label{statistical_weights} \hspace{1.0cm}
\end{equation}
\end{subequations}

\vskip 5mm

\noindent What Eq. (\ref{statistical_weights}) tells us is that
these statistical weights $\omega^{N\pm q}$ are determined by the
ratio between the fraction of charge transferred $\nu$ and the
maximum transfer charge $q$. Also, combining Eqs. (\ref{zero-order})
and (\ref{statistical_weights}), we can write the statistical
weights of the anionic/cationic systems $\omega^{N\pm q}$ in terms
of the central weight $\omega^N$ and the ratio $\nu/q$. Thus, we
finally obtain

\begin{equation}
\omega^{{N \pm q}}(\nu/q)=\frac{1 \pm \nu/q - \omega^{N}(\nu/q)}{2},
\label{statistical_weights_final}
\end{equation}
\vskip 3mm

\noindent where the coefficients $\omega^{{N \pm q}}$ are bounded
(Eq. (\ref{constraint_statistical_weights})) and must fulfills the
following physical conditions at the points $\nu=0$ and $\nu = \pm
q$:

\begin{subequations}
\begin{equation}
\omega^{{N}}(0)=1, \hspace{1.0cm} \omega^{{N \pm q}}(0)=0,
\label{constraint1}
\end{equation}

\noindent as well as

\begin{equation}
\omega^{{N \pm q}}(\pm 1)=1, \hspace{1.0cm} \omega^{{N}}(\pm 1)=\;
\label{constraint2} \omega^{{N \pm q}}(\mp 1)\; =\; 0.
\end{equation}
\end{subequations}

\vskip 3mm

The density matrix $D$ for this three-state model (Eq. (\ref{D1}))
along with the statistical coefficients defined by the relation
(\ref{statistical_weights_final}), fulfilling also Eqs.
(\ref{constraint1}-\ref{constraint2}), represents the more general
quantum state for an interacting molecular domain inside the
molecular frame. Such states can be represented in a $2$-simplex
diagram (equilateral triangle), at whose vertices are attached the
quantum states for integer particle number, $N$, $N+q$, and $N-q$,
while the quantum states associated with non-integer particle number
lie in its interior (Figure 1).



\vskip 5mm

\noindent {\large {\bf 3 Discussion, remarks and conclusions}}

\vskip 3mm

The solutions for the statistical weights $\omega^{{M}}$ clearly
show the acceptor/donor character of the domain $\Omega$ for all
accessible quantum states as a function of the fundamental physical
magnitudes  $\nu$ and $q$.

The solutions expressed by Eq. (\ref{statistical_weights_final}),
determine the accessible quantum states for the domain. They are
completely defined by each point inside the $2$-simplex (Figure 1),
i.e., points that fulfill the condition $0 \le \omega^{M} \le 1$.
The coordinates of points inside the $2$-simplex are given by
$\left( \nu/q ,\omega^N \right)$. Thus, all points over a horizontal
line share the same value of $\omega^N$, as can be seen from Figure
1. Then, for an arbitrary point in any horizontal line, the
corresponding central weight $\omega^N$ can be evaluated from the
value that it possesses over one of the edges of the $2$-simplex
diagram, namely, $\gamma^{\pm}$. These edges represent the maximum
value of charge fraction, for a given statistical weight $\omega^N$,
that the domain can accept or donate, depending on whether we are in
the acceptor region (right, $+$ sign) or the donor region (left, $-$
sign), respectively. These values turn out to be $1 \mp \nu_0/q$,
which correspond to the value of $\omega^N$ according to the
specific value of the transfer rate $\nu_0$. This new physical
magnitude $\nu_0$ corresponds to the limit of the charge fraction
(reference) over the domain for each state that lies in the same
horizontal line. Then, according to Eq.
(\ref{statistical_weights_final}), the weights of the
anionic/cationic system are given by

\vskip 3mm

\[
\omega^{{N \pm q}}(\nu/q; \nu_{0}/q)=\frac{\pm \nu + \nu_{0}}{2q},
\hspace{1.5cm} for \hspace{0.5cm} \Delta^{+}
\]
\[
\omega^{{N \pm q}}(\nu/q; \nu_{0}/q)=\frac{\pm\nu - \nu_{0}}{2q},
\hspace{1.5cm} for \hspace{0.5cm} \Delta^{-}
\]

\vskip 3mm

\noindent where $\Delta^{+}$ and $\Delta^{-}$ stand for the right
and left-hand side of the $2$-simplex in Figure 1 according to the
sign of $\nu$ and $\nu_{0}$. Therefore, replacing these coefficients
in Eq. (\ref{D2}), the quantum state $D$ reads

\begin{equation}
D= \left(1 \mp \frac{\nu_{0}}{q}\right)\; {^{N}\hspace{-0.1cm}D}_{0} +\;
\left(\frac{\nu \pm \nu_{0}}{2q} \right)\; {^{N+q}D}_{0}\; +\;
\left(\frac{-\nu \pm \nu_{0}}{2q} \right)\; {^{N-q}D}_{0}, \label{D3}
\end{equation}

\vskip 5mm

\noindent where the upper and lower signs in Eq. (\ref{D3}) stand
for the accessible states in the regions $\Delta^{\pm}$ shown in
Figure 1 which correspond to the physical acceptor/donor character
of the molecular domain, respectively. It is easy to check that
$Tr(D)=1$, as expected due to its statistical interpretation where
$Tr$ means the mathematical trace symbol \cite{Lowdin}.


The molecular domains for which the convexity property does not hold
as it is the case for an isolated molecule as explained above, no
variational principle can be used to calculate the energy of an
arbitrary state except for those of the limiting edges $\gamma^{\pm}$
\cite{Boch_Rial_JCP_2012}. However, once known the form of the
density matrix $D$, the energy of the domains can be
straightforwardly obtained by applying the mathematical trace
formula ${\cal E^{\cal N}}=Tr({\cal H}D)$ \cite{Lowdin}. Thus, we
obtain for the energy

\begin{equation}
{\cal E}^{\cal N}= \left(1 \mp \frac{\nu_{0}}{q}\right)\; {{\cal
E}_{0}^{N}} +\; \left(\frac{\nu \pm \nu_{0}}{2q} \right)\; {{\cal
E}_{0}^{N+q}}\; +\; \left(\frac{-\nu \pm \nu_{0}}{2q} \right)\;
{{\cal E}_{0}^{N-q}}. \label{energy1}
\end{equation}

\vskip 5mm

To properly understand the physical significance of the states
within each region $\Delta^{\pm}$ inside the $2$-simplex, let us
take some adequate limits for $\nu$ and $\nu_{0}$ in Eqs. (\ref{D3})
and (\ref{energy1}) which will be useful for the discussion. It is
important, at this point, to emphasize the physical significance of
the charge transfer fractions $\nu$ and $\nu_{0}$: the former
corresponds to the charge of fraction transferred in a
\textit{non-equilibrium} state (states inside of the $2$-simplex);
whereas the latter corresponds to the ground states that lie in the
edges of the diagram, being these the reference for the transference
fraction, for a given $\omega^N$, as will be shown in advance.

The limit $\nu \rightarrow \nu_{0}$ means that the system approaches
to the physical limit of the charge transfer determined by $\nu_{0}$
for a given $\omega^N$. Thus the corresponding state with the
associated energy become

\vskip 3mm

\begin{equation}
D= \left(1 \mp \frac{\nu_{0}}{q}\right) {^{N}\hspace{-0.1cm}D}_{0}\; \pm\; \frac{
\nu_{0}}{q}\; {^{N \pm q}D}_{0}, \label{D4}
\end{equation}

\noindent and

\vskip 3mm

\begin{equation}
{\cal E}_{0}^{\cal N}= \left(1 \mp \frac{\nu_{0}}{q}\right) {{\cal
E}_{0}^{N}}\; \pm\; \frac{ \nu_{0}}{q}\; {{\cal E}_{0}^{N \pm q}}. \label{energy0}
\end{equation}

\vskip 3mm

\noindent The quantum state obtained by means of this limit (Eq.
(\ref{D4})) is in accordance with the two-ground state model shown
in Ref. \cite{Boch_Rial_JCP_2012}. Such a two-state model accounts
for the convexity of the energy in atomic and molecular systems
driven by the Coulomb interaction \cite{Parr_Yang_book}.

\vskip 3mm

Another situation that provides interesting physical insights about
the states and the energy for the domains result from the limit $\nu
\rightarrow 0$ (with $\nu_0 \ne 0$), obtaining thus

\begin{equation}
D= \left(1 \mp \frac{\nu_{0}}{q}\right) {^{N}\hspace{-0.1cm}D}_{0}\; \pm\;
\frac{\nu_{0}}{2q} \; {^{N+q}D}_{0}\; \pm\;
\frac{\nu_{0}}{2q} \; {^{N-q}D}_{0}. \label{D5}
\end{equation}

\vskip 3mm

\noindent Such a limit represents the state in which the domain is
neither acceptor nor donor, i.e., is a neutral entity with zero net
charge. Graphically these states lie on the vertical axis of the
$2$-simplex. We can notice that the state shown in Eq. (\ref{D5}) is
only a function of the reference charge fraction transferred at the
ground state $\nu_0$. As it is expected by symmetry arguments, both
$^{N+q}D_{0}$ and $^{N-q}D_{0}$ contribute to the density matrix $D$
in Eq. (\ref{D5}) with the same weight.

\vskip 3mm

From Eq. (\ref{D5}), we can continue exploring the accessible
quantum states taking another interesting limit: $\omega^N
\rightarrow 0$. Within this limit, we have that $\nu_0 = \pm q$. The
quantum state obtained after taking this limit is such one that is
attached to the origin of the $2$-simplex in Figure 1 and is given
by

\vskip 3mm

\begin{equation}
D= \frac{1}{2} \left({^{N+q}D}_{0}\; +\; {^{N-q}D}_{0}  \right)
\end{equation}

\vskip 3mm

\noindent which is nothing but the arithmetic mean of the two ionic
states, an accessible state with no contribution of the neutral
state.

\vskip 3mm

\noindent Finally, let us inspect the accessible states attached at
the vertices of the $2$-simplex. These states corresponds to the
pure states ${^{N}\hspace{-0.1cm}D}_{0}$, ${^{N+q}D}_{0}$ and
${^{N-q}D}_{0}$, whose coordinates are defined by
$\nu=\nu_0=0,\hspace{0.2cm} \omega^N=1$ (upper vertex),
$\nu=\nu_0=q, \hspace{0.2cm}\omega^N=0$ (right bottom vertex) and
$\nu=\nu_0=-q,\hspace{0.2cm}\omega^N=0$ (left bottom vertex),
respectively.

\vskip 3mm

\noindent Once we have analyzed the accessible states it is now the
time for the energies. Thus, it is of important concern to compare
the values of the energies for every point $\bullet$ inside Figure 1
${\cal E}_{\Delta^{\pm}}^{\cal N}(\bullet)$ (cf. Eq. \ref{energy1}),
which are representative of non-equilibrium states (acceptor and
donor states), with those of the corresponding ground states
described by the edges $\gamma^{\pm}$, ${\cal E}_{0}^{\cal
N}(\gamma^{\pm})$. To perform this task we will follow the same
procedure used in the calculation of the ionization potential and
electron affinity in a molecular system, i.e., evaluating the energy
difference between both acceptor/donor states and the corresponding
ground states \cite{Szabo_book}, following the definition

\begin{equation}
I^{\pm} =\; \pm \left({\cal E}_{0}^{N \mp 1} - {\cal
E}_{0}^{N}\right) \label{IP_EA}
> 0,
\end{equation}

\vskip 3mm

\noindent where $I^{+}\equiv I$ stands for the ionization potential,
while $I^{-}\equiv A$ is the electron affinity. Therefore, in the
sense of Eq. (\ref{IP_EA}) and by subtraction of Eqs.
(\ref{energy1}) and (\ref{energy0}) we obtain

\begin{equation}
\Delta H^{\pm}\; :=\; \pm \left(\; {\cal E}_{\Delta^{\pm}}^{\cal
N}(\bullet) - {\cal E}_{0}^{\cal N}(\gamma^{\pm})\right).
\end{equation}

\vskip 3mm

\noindent The energy difference $\Delta H^{\pm}$ can also be
written, explicitly, as

\begin{equation}
\Delta H^{\pm}\; =\; \mp
\left(\frac{\nu-\nu_{0}}{2q}\right)\left(   A^{q}+I^{q}  \right),
\end{equation}

\vskip 3mm

\noindent where $A^{q}$ and $I^{q}$ stand for the $q$-th electron
affinity and $q$-ionization potential defined by $A^{q}={\cal
E}_{0}^{N}-{\cal E}_{0}^{N+q}$ and $I^{q}={\cal E}_{0}^{N-q}-{\cal
E}_{0}^{N}$, respectively. It is experimentally known that $A^{q}$
is, in general, a negative magnitude while $I^{q}$ is always
positive and $|A^{q}|$ < $I^{q}$ ($\abs{.}$ means absolute value).
Hence, the accessible acceptor states have greater energy than the
ground states, while the accessible donor states have a lower one.
Another way to write this measure is

\vskip 3mm

\begin{equation}
\Delta H^{\pm}=\; \pm \left(\frac{\nu-\nu_{0}}{q}\right) \mu_{0}
\geq 0, \label{Delta H mu0}
\end{equation}

\vskip 3mm

\noindent where $\mu_{0}=-\left(\frac{A^{q}+I^{q}}{2}\right)<0$ is
the generalized definition of the {\it chemical potential}
\cite{Geerlings,Parr_Yang_book}. From Eq. (\ref{Delta H mu0}), we
can see that the sign for $\Delta H^{\pm}$ is defined only by the
relative  charge transfer fraction $\nu-\nu_0$. We will discuss this
matter in the following paragraph.

\vskip 3mm

The physical meaning of Eq. (\ref{Delta H mu0}) is that all
accessible acceptor/donor states, i.e., those lying inside the
$\Delta^{\pm}$ have a higher/lower energy than the ground states
defined for the edges $\gamma^{\pm}$, as shown above taking the
appropriate limits. This result is analogous of that of the
two-state model about the stability of the acceptor/donor systems
\cite{PPLB,Boch_Rial_JCP_2012}, reflecting the empirical convexity
of the energy and its monotonically decreasing behavior
\cite{Parr_Yang_book}, namely, the $N \pm q$-state is lower/higher
in energy than the neutral $N$-state, respectively. See Figure 1 in
Ref. \cite{PPLB} and Figure 4.1 in Ref. \cite{Parr_Yang_book}.
Moreover, it is possible to show the inequality

\begin{equation}
{\cal E}_{\Delta^{-}}^{\cal N}(\bullet) > {\cal E}_{\Delta^{+}}^{\cal N}(\bullet),
\end{equation}

\noindent whose meaning is that the following: if we move from the
left to the right inside the $2$-simplex along a horizontal line
with a constant $\omega^N$, varying the value of $\nu/q$ within the
interval $-1< \nu/q < 1$, we obtain that the energy decreases from
the {\it donor} edge $\gamma^-$ towards the {\it acceptor} edge
$\gamma^+$. Therefore, the energy follows the following trend

\vskip 3mm

\begin{equation}
{\cal E}_{0}^{\cal N}(\gamma^{-}) > {\cal E}_{\Delta^{-}}^{\cal
N}(\bullet) > {\cal E}_{\Delta^{+}}^{\cal N}(\bullet) > {\cal
E}_{0}^{\cal N}(\gamma^{+}). \label{trend}
\end{equation}

\vskip 3mm

\noindent This result is analogous to that of figure in Ref. \cite{PPLB}.

On the other hand, we can also perform a variation of the energy
along a vertical line, as shown in Figure 1, namely, variations of
$\omega^{N}$ with $\nu/q$ constant. In this way, we can define a new
quantity $\Delta U^{\pm}$ given by

\begin{equation}
\Delta U^{\pm}:=\mathcal{E}^{\mathcal{N}}_{\Delta \pm}\left( \nu'_0
\right) \label{Delta U} -
\mathcal{E}_{\Delta\pm}^{\mathcal{N}}\left( \nu_0 \right).
\end{equation}

\vskip 3mm

\noindent  In the above relation, it is important to notice that
$\mathcal{E}_{\Delta \pm}^{\mathcal{N}}\left( \nu_0 \right)$ and
$\mathcal{E}_{\Delta \pm}^{\mathcal{N}}\left( \nu'_0 \right)$ share
the same value of $\nu$. However, $\nu_0$ will be higher or smaller
than $\nu'_0$ depending on whether we are in the acceptor zone
$\Delta^{+}$ or in the donor zone $\Delta^{-}$, respectively. Thus,
using Eqs. (\ref{energy0}) and (\ref{energy1}) in (\ref{Delta U}),
we get,

\begin{equation}
\Delta U^{\pm}= \pm \left( \dfrac{\nu_0-\nu'_0}{q}\right)\left(
\mathcal{E}^{N}_0 - \mathcal{\bar{E}}_0 \right) , \label{Delta U 2}
\end{equation}

\vskip 3mm

\noindent where $\mathcal{\bar{E}}_0$ means the arithmetic average
between the acceptor and donor energies $\mathcal{E}_0^{N+q}$ and
$\mathcal{E}_0^{N-q}$, i.e.,
$\mathcal{\bar{E}}_0=\frac{\mathcal{E}_0^{N+q}+\mathcal{E}_0^{N-q}}{2}$.
At this point, it is worth mentioning that the arithmetic average
$\mathcal{\bar{E}}_0$ is always higher than the energy of the
neutral system $\mathcal{E}^N_0$, i.e.,

\begin{equation}
\mathcal{\bar{E}}_0>\mathcal{E}^N_0, \label{average>eN}
\end{equation}

\noindent which easily follows from $I^q>|A^q|$. Then, from Eq.
(\ref{average>eN}), and considering the relation between $\nu_0$ and
$\nu'_0$ depending on the region of the diagram we are in, it is
possible to show that

\begin{equation}
\Delta U^{\pm} \leq 0.
\end{equation}

\noindent Therefore, we can conclude that

\begin{equation}
\mathcal{E}_{\Delta\pm}^{\mathcal{N}}\left( \nu'_0 \right)\leq
\mathcal{E}^{\mathcal{N}}_{\Delta \pm}\left( \nu_0 \right),
\end{equation}

\vskip 3mm

\noindent which means that if we compare two different states within
the diagram connected through a vertical line, we have that the
energy of the state below will be higher than the energy of the
state above.

In order to express this result explicitly, let us rewrite the
energy difference between the neutral system  $\mathcal{E}^N_0$ and
the arithmetic average $\mathcal{\bar{E}}_0$ as

\begin{equation}
\mathcal{E}_0^{N}-\mathcal{\bar{E}}_0 = \dfrac{I^q-A^q}{2}.
\end{equation}

\vskip 3mm

\noindent Then, using the definition of the molecular hardness,
which is given by \cite{Geerlings}

\begin{equation}
\eta_0=\dfrac{I^q-A^q}{2},
\end{equation}
\vskip 3mm
\noindent we finally can recast the vertical energy variation $\Delta U ^{\pm}$ as

\begin{equation}
\Delta U^{\pm}=\pm \left( \dfrac{\nu'_0 - \nu_0}{q}\right)\eta_0.
\label{Delta U eta0}
\end{equation}

\vskip 3mm

\noindent Thus, while the energy difference along a horizontal line
with $\omega^N$ constant is related to the chemical potential
$\mu_0$, a vertical energy difference is related to the molecular
hardness $\eta_0$.


Let us finally discuss the variation of the energy of the molecular
domain shown in Figure 1. These variations were calculated among the
accessible quantum states of the domain using two different
quantities, namely $\Delta H^{\pm}$ and $\Delta U^{\pm}$,
representing both two different types of electronic processes, and
expressed by Eqs. (\ref{Delta H mu0}) and (\ref{Delta U eta0}),
respectively.

The energy variation given by $\Delta H^{\pm}$ keeps the coefficient
$\omega^{N}$ of the neutral system fixed in the convex expansion of
the states (horizontal line in Figure 1); this variation is driven
by the relative difference of the charge transfer $\nu-\nu_{0}$
which is proportional to the chemical potential of the neutral
system $\mu_0$. Thus, such a horizontal line, connecting both edges
$\gamma^-$ and $\gamma^+$ of the $2$-simplex follows the trend given
in (\ref{trend}). In this way, the statistical weights of the
anionic/cationic systems, $\omega^{\pm}$, change from the $\gamma^-$
($\omega^{N+q}=0$) to $\gamma^+$ ($\omega^{N-q}=0$). These changes
are in accordance with which are shown in Ref.  \cite{Boch_Rial_JCP_2012}
for the two-state model. Hence, the above considerations allow us to
conclude that, as long as the charge fraction $\nu$ increases, the
molecular domain stabilizes.

The energy variation of the states along a vertical line $\Delta
U^{\pm}$ (Eq. (\ref{Delta U eta0})) shares the charge fraction $\nu$
for both initial and final states. Thus, once the energy difference
is evaluated, it is shown that it depends only on the difference
$\abs{\nu'_0-\nu_0}$ between the reference charge fraction
transferred in the ground state. Hence, the proportionality factor
is nothing but the chemical hardness $\eta_0$ of the molecular
domain. In this case, the energy difference $\Delta U^{\pm}$ depends
on the net charge fraction $\nu$ only indirectly: $\nu$ only
indicates {\it how far} is the vertical line to the edges
$\gamma^{\pm}$. Also, due to the constancy of the net charge
fraction $\nu$ and the shape of the domain during the energy
variation along a vertical line, the domain does not have the
possibility of polarizing its electronic distribution. Therefore,
with the above in mind, it is reasonable that the proportionality
factor, in this case, be the chemical hardness according to its
physical meaning given in Ref. \cite{Geerlings}.

These final considerations open a question about the proper
definition of this kind of descriptors supported by the derivatives
of the energy, namely,  chemical potential, hardness among others,
which collides with the discontinuities at the integer number of
particles \cite{Parr_Yang_book,Geerlings,Wasserman}, and
consequently merits to consider a proper definition for these
magnitudes. We plan to consider these topics in forthcoming
developments.

\begin{figure}
  \includegraphics[width=\linewidth]{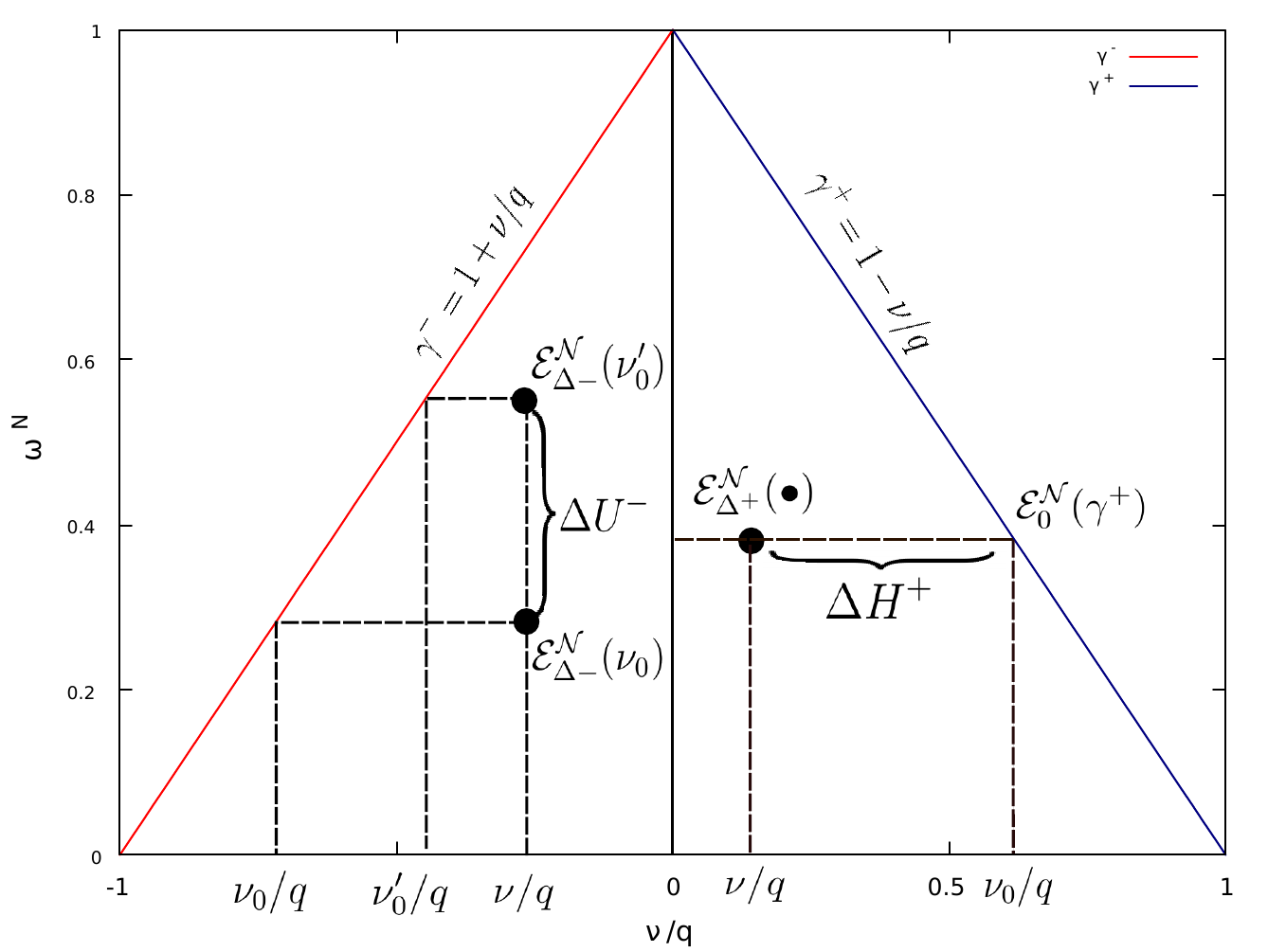}
  \caption{$\omega^{{N}}$ versus $\nu/q$. The $2$-simplex diagram (equilateral triangle) corresponds to a mathematical $2$-dimensional region into which coefficients $0 \le \omega^{{N \pm q}}(\nu) \le 1$ in the convex expansion of the Density Matrix $D$ are bounded according to Eq. (\ref{constraint_statistical_weights}). Subtriangle regions $\Delta^{\pm}$ represent the acceptor ($\nu > 0$) and donor ($\nu < 0$) accessible states, respectively}
\end{figure}

\newpage

\vskip 10mm

\noindent {\bf Acknowledgments}

\noindent This report has been financially supported by Projects
20020130100226BA (Universidad de Buenos Aires) and PIP No.
11220090100061 (Consejo Nacional de Investigaciones Cient\'{\i}ficas
y T\'ecnicas, Rep\'ublica Argentina). BM acknowledges project N° DI-06-23/PASAN of the Vicerrector\'ia de Investigaci\'on y Doctorado, Universidad Andres Bello (Chile) and was realized in the context of ``Concurso Pasant\'ias de Investigaci\'on en el
Extranjero para Tesistas de Doctorado 2023'', during spring of 2023 in the city of Buenos Aires, Argentina.

\newpage

\section*{References}



\newpage


\end{document}